\documentclass[pre,twocolumn,aps,eqsecnum]{revtex4}
\usepackage{amsmath,bm,epsfig}
\newcommand{\B}[1]{{\bm{#1}}}

\begin{document}

\title{Nonequilibrium Thermodynamics of Amorphous Materials III:\\
 Shear-Transformation-Zone Plasticity}
\author{Eran Bouchbinder}
\affiliation{Racah Institute of Physics, Hebrew University of Jerusalem, Jerusalem 91904, Israel}
\author{J. S. Langer}
\affiliation{Dept. of Physics, University of California, Santa Barbara, CA 93106-9530}

\begin{abstract}
We use the internal-variable, effective-temperature thermodynamics developed in two preceding papers to reformulate the shear-transformation-zone (STZ) theory of amorphous plasticity. As required by the preceding analysis, we make explicit approximations for the energy and entropy of the STZ internal degrees of freedom. We then show that the second law of thermodynamics constrains the STZ transition rates to have an Eyring form as a function of the effective temperature. Finally, we derive an equation of motion for the effective temperature for the case of STZ dynamics.
\end{abstract}

\maketitle

\section{Introduction}
\label{intro}

Understanding the irreversible deformation of amorphous systems remains a major challenge in nonequilibrium statistical physics and materials science \cite{01LN, 06MR}. Systems of interest include noncrystalline solids below or near their glass temperatures, dense granular materials, and various kinds of soft materials such as foams, colloids, and the like. An ongoing effort to develop a dynamical theory of such systems has been based on the shear-transformation-zone (STZ) model of \cite{FL98}. Recent work has extended the original model to include an effective disorder temperature as an essential ingredient \cite{JSL04,BLP07I,BLP07II,JSL-MANNING-TEFF-07, EB-TEFF-PRE08, JSL-STZ-PRE08,SHI-FALK-SHEARBANDS-07,MANNINGetal-SHEARBANDS-08}.

Our main goal in this paper is to develop an STZ theory that is consistent with the internal-variable, effective-temperature thermodynamics described in two preceding papers \cite{EB-JSL-09-I, EB-JSL-09-II}. In \cite{EB-JSL-09-I} we focused on the role of internal state variables in determining the nonequilibrium dynamics of amorphous, not necessarily glassy, systems. We used the statistical interpretation of the first and second laws of thermodynamics to obtain equations of motion for the internal variables, and we emphasized the need to understand how both energy and entropy are shared between the internal variables and other degrees of freedom.

In \cite{EB-JSL-09-II} we extended this development to include an effective disorder temperature. Our basic premise in that paper was that the slow configurational degrees of freedom of such materials are only weakly coupled to the fast kinetic/vibrational degrees of freedom, and therefore that these two subsystems can be described by different temperatures during deformation. Using the tools of nonequilibrium statistical thermodynamics, we derived a general form for the equation of motion for the effective temperature, and obtained a set of second-law constraints on the thermomechanical equations of motion for such systems.

We start here in Sec. \ref{thermo} by summarizing the major results of \cite{EB-JSL-09-I, EB-JSL-09-II} in a form appropriate for the STZ analysis. In Sec. \ref{STZ1}, we introduce the STZ degrees of freedom as thermodynamically well defined internal state variables with associated energies and entropies. We then deduce specific forms for the STZ equations of motion based on the thermodynamic analysis.  Our most important departure from earlier versions of the theory is that the STZ transition rates are now required to have an Eyring form as a function of the effective temperature rather than the reservoir temperature. In Sec. \ref{STZ2}, we discuss the noise strength that determines the STZ annihilation and creation rates, and we derive an equation of motion for the effective temperature. Section \ref{STZsummary} contains a summary of the STZ equations. We conclude in Sec. \ref{conclusions} with remarks about the significance and limitations of this theory.

\section{Thermodynamic Constraints}
\label{thermo}

We consider the deformation of an amorphous material in contact with a thermal reservoir at temperature $\theta_R$. We assume that $\theta_R$ is either below or not too far above the glass temperature $\theta_g$, so that the two-temperature theory developed in \cite{EB-JSL-09-II} is applicable. We express temperatures in units of energy, and set Boltzmann's constant $k_B$ equal to unity. For simplicity, we assume from the beginning that the system is spatially uniform and that it undergoes only volume-conserving, pure-shear deformations.

The total, extensive, internal energy of this system, including a thermal reservoir, is
\begin{equation}
\label{Ueqn}
U_{tot} \cong U_C(S_C,{\B E}_{el},\{\Lambda_{\alpha}\}) + U_K(S_K) + U_R(S_R) \ ,
\end{equation}
where $U_C$ and $U_K$, respectively, are the configurational and kinetic/vibrational internal energies, $S_C$ and $S_K$ are the respective entropies, and $\{\Lambda_{\alpha}\}$ denotes a set of internal state variables, soon to be identified as the STZ variables. $U_R$ is the energy of the thermal reservoir, which we assume to be strongly coupled to the kinetic/vibrational subsystem. ${\B E}_{el}$ is a deviatoric (traceless, symmetric), elastic shear strain. Note that our assumption of volume-conserving, pure-shear deformation allows us to omit any volume dependence in $U_K$, cf. Eq. (3.1) in \cite{EB-JSL-09-II}.

The effective temperature $\chi$ and the kinetic/vibrational temperature $\theta$ are
\begin{eqnarray}
\label{theta-U}
\chi = \left({\partial U_C \over \partial S_C}\right)_{{\B E}_{el}},~~\theta = \left({\partial U_K \over \partial S_K}\right)_{V_{el}} \ .
\end{eqnarray}
We assume that $\theta\!\approx\!\theta_R$, i.e. that the kinetic/vibrational subsystem is always in equilibrium with the thermal reservoir. The shear stress acting on the configurational subsystem is
\begin{equation}
\label{s-U}
V\,{\B s}_C = \left({\partial U_C\over \partial {\B E}_{el}}\right)_{S_C,\{\Lambda_\alpha\}},
\end{equation}
where $V$ is the fixed total volume. As explained in \cite{EB-JSL-09-II}, the kinetic/vibrational subsystem has no shear modulus, but it can support a viscous  stress in the presence of shear flow.  For further simplicity, we assume that the kinetic/vibrational viscosity vanishes.

The total entropy is
\begin{equation}
\label{Seqn}
S_{tot} \simeq S_C(U_C,{\B E}_{el},\{\Lambda_{\alpha}\})+ S_K(U_K) + S_R(U_R) \ .
\end{equation}
The expression for any one of these three entropies can be inverted to obtain the corresponding internal energy function in Eq. (\ref{Ueqn}), or {\it vice versa}.

Without further loss of generality, we specialize to the case of pure, planar shear oriented along fixed axes, say $x$ and $y$, and define $s_C \equiv s_{C,xx} = -\,s_{C,yy}$.  We assume (for small elastic deformations) that the rate of deformation tensor is the sum of elastic and inelastic parts,  $\B D = \B D_{el}+\B D_{in}$, where $\B D_{el}=\dot{\B E}_{el}$, and we define $D_{in} \equiv D_{in,xx} = - D_{in,yy}$.  All other elements of these deviatoric tensors vanish; thus, for example, the rate of inelastic work done by the shear stress is ${\B s}_C:{\B D}_{in}= 2\,s_C\,D_{in}$.

The analysis in \cite{EB-JSL-09-II} produced an equation of motion for the effective temperature that is basically a statement of the first law of thermodynamics, i.e. a heat-flow equation.  For the present case, this equation has the form
\begin{equation}
\label{chi_EOM}
C_V^{e\!f\!f}\,\dot \chi =
{\cal W}_C(s_C,\{\dot\Lambda_\alpha\}) +  A(\chi,\theta)\,\left(1-{\chi\over \theta}\right).
\end{equation}
Here
\begin{equation}
C_V^{e\!f\!f}\,\dot \chi = \chi \dot S_C
\end{equation}
is the time rate of change of the heat of configurational disorder, and $C_V^{e\!f\!f}$ is an effective (extensive) heat capacity at constant volume. As in the preceding papers, the non-negative dissipation rate ${\cal W}_C$ -- the difference between the rate at which inelastic work is being done on the configurational subsystem and the rate at which energy is being stored in the internal degrees of freedom -- is
\begin{equation}
\label{calWdef}
{\cal W}_C (s_C,\{\dot\Lambda_\alpha\}) = 2\,V\,s_C\, D_{in}
-\sum_{\alpha} \left({\partial U_C\over \partial \Lambda_{\alpha}}\right)_{S_C,{\B E}_{el}}\dot \Lambda_{\alpha} \ge 0.
\end{equation}
Non-negativity of ${\cal W}_C$ is an important second-law constraint that plays a central role in the analysis to follow.

The second term on the right-hand side of Eq.(\ref{chi_EOM}),
\begin{equation}
\label{Qdef}
A(\chi,\theta)\,\left(1-{\chi\over \theta}\right) \equiv Q
\end{equation}
is the rate at which heat is flowing into the configurational subsystem.  Here, $A(\chi,\theta)$ is a non-negative thermal transport coefficient that, as will be seen, depends on other dynamical variables in addition to $\chi$ and $\theta$.

\section{STZ Equations of Motion}
\label{STZ1}

The basic assumptions of the STZ theory have been described in \cite{JSL-STZ-PRE08}. To the extent possible, the following discussion follows the steps outlined in that paper.

The main idea is that deformation of amorphous materials occurs via localized molecular rearrangements that take place at shear transformation zones (STZ's).  The STZ's are created and annihilated either by thermal fluctuations or by noise generated by the deformation itself.  They are rare, ephemeral fluctuations that are especially important for irreversible deformations because they make stress-driven transitions between two, energetically almost degenerate orientations.  Thus, the STZ's are two-state systems. There is nothing arbitrary about this two-state picture. The STZ's have the special property of being able to shift between one orientation and another in response to a shear stress. Sites with this property are already statistically unlikely, and higher-order degeneracies are statistically negligible.

The difference between what we are doing here and the analysis presented in \cite{JSL-STZ-PRE08} is that now, on the basis of \cite{EB-JSL-09-I, EB-JSL-09-II}, we insist on a proper thermodynamic description of the STZ's as internal degrees of freedom.  Such a description requires a specific STZ model.  To construct any such model, we must make physical assumptions that may need to be modified in later applications.  In particular, as in the earlier work, we  assume that there is just a single kind of STZ, with a single characteristic formation energy $e_Z$, and a single mechanism for making transitions between the two orientational states.

For additional simplicity, we go back to the original version of the theory \cite{FL98} in which the STZ's occur only with  orientations either ``$+$'' or ``$-$'' with respect to the shear direction. A procedure for averaging over STZ orientations and constructing a properly invariant tensorial version of the theory was presented in \cite{JSL-STZ-PRE08}. That procedure works just as well for the present analysis, but seems unnecessarily complex for present purposes. The internal state variables are the extensive numbers of STZ's in these two different states, $N_+$ and $N_-$.  As usual, define
\begin{equation}
\Lambda \equiv {N_+ + N_-\over N},~~~~ m \equiv {N_+ - N_-\over N_+ + N_-}.
\end{equation}
Thus, the set of internal state variables $\{\Lambda_\alpha\}$ reduces to $\{\Lambda, m\}$.

Our arguments in \cite{EB-JSL-09-I,EB-JSL-09-II} tell us that we must include the entropy associated with the internal variables $\Lambda$ and $m$ in this analysis.  If we take the two-state model literally, then we compute this entropy by counting the number of ways in which we can distribute $N_+$ ``$+$'' zones and $N_-$ ``$-$'' zones among, say, $N$ available sites in the system. This number is
\begin{equation}
\exp\,(S_Z) = {N!\over N_+!\,N_-!\,(N - N_+ - N_-)!},
\end{equation}
which, after use of Stirling's approximation, reduces to
\begin{equation}
\label{SZ}
{S_Z(\Lambda,m)\over N} \approx - \Lambda\,\ln\Lambda - (1 - \Lambda)\,\ln(1- \Lambda)  + \Lambda \,S_0(m),
\end{equation}
where
\begin{equation}
\label{S0def}
S_0(m) = \ln 2 - {1\over 2}\,(1+m)\,\ln(1+m) - {1\over 2}\, (1-m)\,\ln(1-m).
\end{equation}

To use this formula, write
\begin{equation}
\label{SCdef}
S_C = S_Z(\Lambda,m) + S_1(U_1),
\end{equation}
where $S_1$ and $U_1$, respectively, are the entropy and energy of all the degrees of freedom of the configurational subsystem apart from those attributable to the STZ's. Accordingly,
\begin{eqnarray}
\label{UCdef}
\nonumber
&&U_C(S_C, \B E_{el}, \Lambda, m) = N\,\Lambda\,e_Z + U_1(S_1,{\B E}_{el}) \\
\cr &&= N\,\Lambda\,e_Z + U_1\big[S_C-S_Z(\Lambda,m),{\B E}_{el}\bigr],
\end{eqnarray}
where $e_Z$ is the formation energy of an STZ.  Equations (\ref{SCdef}) and (\ref{UCdef})  are equivalent to each other if we write $U_1 = U_C - N\,\Lambda\,e_Z$ in Eq. (\ref{SCdef}).

In terms of these STZ variables, the inequality in Eq. (\ref{calWdef}) becomes
\begin{eqnarray}
\label{calWdef2}
\nonumber
&&{\cal W}_C(s_C,\{\dot\Lambda_\alpha\}) \to {\cal W}_C(s_C,\dot\Lambda, \dot m)\cr \\&&\nonumber = 2\,V\,s_C\, D_{in}  - N\,\left[e_Z + \chi\,\ln\,\left({\Lambda\over 1 - \Lambda}\right) -\,\chi\,S_0(m)\right] \,\dot \Lambda \cr\\&& + N\,\chi\, \Lambda\,{dS_0\over dm}\,\dot m \ge 0.
\end{eqnarray}

To make further progress, go back to the original STZ equations of motion for the $N_{\pm}$
\begin{equation}
\label{dotNpm}
\tau_0\,\dot N_{\pm} = R(\pm s_C)\,N_{\mp} - R(\mp s_C)\,N_{\pm} + \tilde \Gamma\,\left({1\over 2}\,N^{eq} - N_{\pm}\right).
\end{equation}
Here, $\tau_0$ is a time scale, the factors $R(\pm s_C)/\tau_0$ are the rates at which STZ's switch back and forth between their two orientations, $\tilde \Gamma/\tau_0$ is the rate factor for creation and annihilation of STZ's, and $N^{eq}$ is an as-yet undetermined ``equilibrium'' value for the number of STZ's. The superscript ``eq'' is used here and below to denote steady-state equilibrium. Note that, in Eq. (\ref{dotNpm}), we are assuming that the STZ creation rate is the same for both STZ orientations, independent of the orientational state of the system as a whole.

The deviatoric, inelastic rate of deformation tensor is
\begin{eqnarray}
\nonumber
D_{in} = \frac{v_0}{\tau_0 V}\,\bigl[R(+s_C)\,N_- -\,R(-s_C)\,N_+\bigr],
\end{eqnarray}
where $v_0$ is a molecular-scale volume.  As usual, define
\begin{eqnarray}
\label{CTdef}
\nonumber
{\cal C}(s_C) &\equiv& {1\over 2}\,\bigl[R(+s_C) + R(-s_C)\bigr],\cr\\ {\cal T}(s_C) &\equiv& {R(+s_C) - R(-s_C)\over R(+s_C) + R(-s_C)} \,.
\end{eqnarray}
Then,
\begin{equation}
\label{Ddef}
\tau_0\,D_{in} = {N\,v_0\over V}\,\Lambda\,{\cal C}(s_C)\,\bigl[{\cal T}(s_C) - m\bigr].
\end{equation}
In previous papers, we defined $N\,v_0/V \equiv \epsilon_0$. We will return to this notation in Sec. \ref{STZsummary}. The equations of motion for $\Lambda$ and $m$ are
\begin{equation}
\label{Lambdadot}
\tau_0\,\dot \Lambda = \tilde \Gamma\,\bigl(\Lambda^{eq} - \,\Lambda\bigr),
\end{equation}
where $\Lambda^{eq} = N^{eq}/N$, and
\begin{equation}
\label{mdot}
\tau_0\,\dot m = 2\,{\cal C}(s_C)\,\bigl[{\cal T}(s_C) - m\bigr]- \tilde\Gamma\,m - {\dot\Lambda\over \Lambda}\,m.
\end{equation}

The next step in this analysis is to impose the second-law constraint expressed in Eq. (\ref{calWdef2}). We immediately encounter a difference between the present situation and the one described, for example, by Maugin in \cite{MAUGIN-99}.  Specifically, the inelastic rate of deformation $D_{in}$ appearing in ${\cal W}_C$ is not simply proportional to the time derivatives $\dot \Lambda$ and $\dot m$.  Therefore, we cannot satisfy the inequality in Eq. (\ref{calWdef2}) by identifying the coefficients of those time derivatives as thermodynamic forces associated with energy landscapes, and then requiring that $\Lambda$ and $m$ both relax toward free-energy minima.  In fact, our situation is more interesting.  It is almost certainly typical of open systems in which external work is being done and energy is being dissipated, and where no variational formulation is relevant.

Our strategy is to use Eq. (\ref{mdot}) to evaluate $\dot m$ in Eq. (\ref{calWdef2}), and thereby to write ${\cal W}_C$ as the sum of two terms, one proportional to $\dot\Lambda$, and the other proportional to the stress-dependent quantity ${\cal T}(s_C) - m$. These two terms must individually be non-negative. The inequality in (\ref{calWdef2}) becomes
\begin{eqnarray}
\label{WCineq2}
&&{\tau_0\over N}\,{\cal W}_C(s_C,\dot\Lambda, \dot m) = - \tilde\Gamma\,\chi\,\Lambda\,m\,{dS_0\over dm} \\ \cr && \nonumber - \left[e_Z + \chi\,\ln\,\left({\Lambda\over 1 - \Lambda}\right) -\,\chi\,S_0(m)+ \chi\,m\,{dS_0\over dm}\right] \,\tau_0\,\dot \Lambda\cr \\  &&+ 2\,\Lambda\,{\cal C}(s_C) \bigl[{\cal T}(s_C) - m\bigr]\left(v_0\,s_C + \chi\,{dS_0\over dm}\right) \ge 0 \ .\nonumber
\end{eqnarray}
From Eq. (\ref{S0def}), we know that
\begin{equation}
\label{dS0dm}
{dS_0\over dm} = - {1\over 2}\,\ln\,\left({1+m\over 1-m}\right) = - \tanh^{-1}(m).
\end{equation}
Therefore, the first term in the expression for ${\cal W}_C$ in Eq. (\ref{WCineq2}) is always non-negative, and we can set it aside for the moment.

The second term in Eq. (\ref{WCineq2}) produces a standard, variational, second-law inequality of the form
\begin{equation}
-{\partial F_Z\over \partial \Lambda}\,\dot \Lambda \ge 0,
\end{equation}
where
\begin{equation}
\label{FZ}
F_Z(\Lambda,m) = N\,e_Z\,\Lambda - \chi\,\left[S_Z(\Lambda,m) - m\,{\partial S_Z\over \partial m}\right]
\end{equation}
is a free energy. $\Lambda^{eq}$ in Eq. (\ref{Lambdadot}) must be the value of $\Lambda$ at which
\begin{equation}
 \left({\partial F_Z\over \partial \Lambda}\right)_{\Lambda = \Lambda^{eq}} = 0.
\end{equation}
Therefore,
\begin{equation}
\label{Lambdaeq}
\Lambda^{eq}(\chi,m)= {{\cal Z}^{eq}\over 1 + {\cal Z}^{eq}},
\end{equation}
where
\begin{equation}
\label{Zdef}
 {\cal Z}^{eq}(\chi,m)= \exp\,\left[-{e_Z\over \chi} + S_0(m) - m\,{dS_0\over dm}\right].
\end{equation}
For $\chi \ll e_Z$, we expect $\Lambda^{eq} \approx {\cal Z}^{eq} \ll 1$, which is consistent with the basic idea of a low density of STZ's.  We then obtain the expected Boltzmann factor, $\Lambda^{eq} \approx \exp\,(- e_Z/\chi)$, with a small modification from the $m$-dependent entropy.  The term proportional to $m\,dS_0/dm$ in Eq. (\ref{Zdef}) means that ${\cal Z}^{eq}$ diverges weakly, and $\Lambda^{eq} \to 1$, when $m \to \pm 1$.  However, it is easy to see from the denominator in the equation of motion for $m$, i.e. either Eq.(\ref{mdot1}) or Eq.(\ref{mdot2}) shown below, that  $m \to \pm 1$ is a dynamically inaccessible limit.  Therefore, so long as $e_Z$ is the largest energy scale in the problem -- which has always been the case in prior applications -- the requirement of small $\Lambda$ is satisfied.

The more interesting result comes from the term proportional to ${\cal T}(s_C) - m$ in Eq. (\ref{WCineq2}). That term must be non-negative for all values of the stress $s_C$, i.e.
\begin{equation}
\bigl[{\cal T}(s_C) - m\bigr]\left(v_0\,s_C + \chi\,{dS_0\over dm}\right) \ge 0,
\end{equation}
which means that the two stress-dependent factors, ${\cal T}(s_C) - m$ and $v_0\,s_C + \chi\,dS_0/dm$, must each be monotonically increasing functions of $s_C$ that change sign at the same point for arbitrary values of $m$.  From Eq. (\ref{dS0dm}), we see that this condition can be satisfied only if
\begin{equation}
{\cal T}(s_C) = \tanh\,\left({v_0\,s_C\over \chi}\right),
\end{equation}
which, according to Eq. (\ref{CTdef}), means that
\begin{equation}
\label{Rate}
R(s_C) = R_0(s_C,\chi,\theta)\,\exp\,\left({v_0\,s_C\over \chi}\right),
\end{equation}
where $R_0$ is a symmetric, non-negative function of $s_C$. As indicated, $R_0$ may also depend on the temperatures $\chi$ and $\theta$, because the transitions between STZ orientations are very likely to be thermally activated processes.  Equation (\ref{Rate}) indicates a major difference between the present thermodynamic results and the earlier theories.  In the latter, we started with physical models for the transition rates $R(\pm s_C)$, and then assumed that the dependence of the internal energy on the STZ variables would be consistent with these rates.  Here we start with a known internal energy, and must argue in the other direction to make sure that the rates are consistent with thermodynamics.  In particular, Eq. (\ref{Rate}) tells us that the STZ transition rates must have an Eyring form with the effective temperature $\chi$ rather than the reservoir temperature $\theta$ in the exponent.

\section{Noise Strength and Equation of Motion for $\chi$}
\label{STZ2}

Having used the second law to deduce equations of motion for the STZ variables, our next steps are to go back to the first law in Eq. (\ref{chi_EOM}) and use the expressions for $\dot\Lambda$ and $\dot m$ to compute $\tilde\Gamma$, and then to derive the STZ version of an equation of motion for $\chi$. Both of these steps again require going beyond purely thermodynamic arguments, and making additional physical assumptions.

Equation (\ref{chi_EOM}) now can be expressed explicitly in terms of the internal variables:
\begin{eqnarray}
\label{firstlaw4}
&&C_V^{e\!f\!f}\,\dot \chi - A(\chi,\theta)\,\left(1-{\chi\over \theta}\right) = \chi\,\dot S_C - Q = \\
&&- N\,{\tilde \Gamma\over \tau_0}\,{\partial F_Z\over \partial \Lambda}\,\Bigl(\Lambda^{eq} - \Lambda\Bigr)- N\,{\tilde\Gamma\over\tau_0}\,\chi\,\Lambda\,m\,{dS_0\over dm} \cr \nonumber\\&&\nonumber - {2\,N\,\Lambda\over \tau_0}\,{\cal C}(s_C)\,\bigl[{\cal T}(s_C) - m\bigr]\,\left(v_0\,s_C + \chi\,{dS_0\over dm}\right).
\end{eqnarray}
As in previous STZ papers, we assume that the rate factor $\tilde\Gamma$ is a sum of two independent noise strengths, $\tilde\Gamma = \Gamma(s_C,\chi) + \rho(\theta)$. Here $\Gamma(s_C,\chi)$ is the part of the rate factor determined by mechanically generated noise, and $\rho(\theta)$ is the super-Arrhenius, thermally generated part. We next invoke Pechenik's hypothesis \cite{PECHENIK}, which identifies $\Gamma$ as being proportional to the total rate of heat production per STZ
\begin{equation}
\label{Pechenik}
\chi\,\dot S_C - Q  = {\cal W}_C = {\Gamma(s_C,\chi)\over \tau_0}\,N\,\Lambda\,v_0\,s_0,
\end{equation}
where the proportionality factor $s_0$ has the dimensions of stress.  Inserting this relation into Eq. (\ref{firstlaw4}) and solving for $\tilde\Gamma$, we find
\begin{equation}
\label{Gammadef}
\tilde\Gamma = \Gamma(s_C,\chi) + \rho(\theta)= {\tilde{\cal N}(s_C,\Lambda,m)\over \Delta(\Lambda,m)},
\end{equation}
where
\begin{eqnarray}
\nonumber
&&\tilde {\cal N}(s_C,\Lambda,m) =  \rho(\theta)\,v_0\,s_0\cr\\ && + 2\,{\cal C}(s_C)\,\bigl[{\cal T}(s_C) - m\bigr]\left(v_0\,s_C + \chi\,{dS_0\over dm}\right)
\end{eqnarray}
and
\begin{equation}
\Delta(\Lambda,m) = v_0\,s_0 + {\partial F_Z\over \partial \Lambda}\,\left({\Lambda^{eq}\over \Lambda} - 1\right)+ m\,\chi\,{dS_0\over dm}.
\end{equation}
The equation of motion for $m$, Eq. (\ref{mdot}), becomes
\begin{equation}
\label{mdot1}
\tau_0\,\dot m = {{\cal M}(s_C,\Lambda,m)\over \Delta(\Lambda,m)},
\end{equation}
where
\begin{eqnarray}
\nonumber
&&{\cal M}(s_C,\Lambda,m)= - \,m\,{\Lambda^{eq}\over \Lambda}\,\rho(\theta)\,s_0\,v_0\cr \\ \nonumber && +\, 2\,{\cal C}(s_C)\,\bigl[{\cal T}(s_C) - m\bigr] \,\Biggl[v_0\,(s_0 - m\,s_C)\cr \\&&  - \left({\Lambda^{eq}\over \Lambda} - 1\right)\,\left(v_0\,m\,s_C + m\,\chi\,{dS_0\over dm} - {\partial F_Z\over\partial \Lambda}\right)\Biggr].~~~~
\end{eqnarray}

At this point, it is useful to distinguish between slow and fast processes, as was done in \cite{BLP07II,JSL-STZ-PRE08}.  The inelastic deformation rate given in Eq. (\ref{Ddef}) contains a factor $\Lambda$, meaning that it is proportional to the density of STZ's and is small.  The equation of motion for $\chi$ will be seen to be similarly slow.  On the other hand, the equations of motion for $\Lambda$ and $m$ contain no such factors $\Lambda$. These internal state variables respond rapidly to changes in their environments.  Therefore, we simplify the analysis by setting $\Lambda = \Lambda^{eq}$, and replacing $m$ by $m^{eq}$, the stationary solution of
\begin{equation}
\label{mdot2}
\tau_0\,\dot m = {2\,{\cal C}(s_C)\,\bigl[{\cal T}(s_C) - m\bigr]\,\bigl(1 - m\,s_C/s_0\bigr) - m\,\rho(\theta)\over 1 + (m\,\chi/v_0\,s_0)\,(dS_0/dm)}.
\end{equation}
This solution is shown explicitly in Eq. (\ref{meq}). These approximations are always valid for steady-state solutions but, as seen in \cite{JSL-STZ-PRE08}, they also work well for transients.

In steady state, and at low temperatures where $\rho(\theta) \approx 0$, Eq. (\ref{mdot2}) exhibits the usual \cite{FL98, BLP07I, EB-TSL-fronts-PRE08} exchange of stability at a yield stress (minimum flow stress) $s_y$ determined implicitly by
\begin{equation}
\label{syield}
s_y\,\tanh\,\left({v_0\,s_y\over \chi_0}\right) = s_0,
\end{equation}
where $\chi_0$ is the steady-state value of $\chi$ in the limit of vanishingly small strain rate. According to Eqs. (\ref{mdot2}) and (\ref{meq}), for $\rho(\theta) = 0$, $m^{eq}$ goes through a maximum value of $\tanh\,(v_0\,s_y/\chi_0)$ at $s_C = s_y$. At that point, Eqs. (\ref{Lambdaeq}) and (\ref{Zdef}) tell us that the condition
\begin{equation}
\Lambda^{eq}(\chi_0,m^{eq}) \approx \exp\,\left(- \,{e_Z - v_0\,s_y\over \chi_0}\right) \ll 1
\end{equation}
requires that $e_Z$ be much larger than $\chi_0$ and $v_0\,s_y$, which, as noted earlier, is generally true.

To complete this development, we need an explicit equation of motion for $\chi$, and again we need to make additional physical assumptions. Use Eqs. (\ref{firstlaw4}) and (\ref{Pechenik}) to write
\begin{equation}
\label{chidot}
 C_V^{e\!f\!f}\,\dot \chi = {\Gamma(s_C,\chi)\over \tau_0}\,N\,\Lambda^{eq}\,v_0\,s_0 + A(\chi,\theta)\,\left( 1 - {\chi\over\theta}\right).
\end{equation}
The thermal transport coefficient $A(\chi,\theta)$ is one of two places in this theory where the weak coupling between the configurational and kinetic/vibrational subsystems must be modeled explicitly.  The other place is the noise strength $\Gamma$ defined in Eq.(\ref{Pechenik}), where we argued that mechanically generated noise contributes additively, along with the thermal noise, in creating configurational disorder.  Similarly, it seems plausible that the overall heat exchange between the two subsystems is enhanced by mechanical noise.  Thus we propose that $A$ have a form similar to that of $\tilde\Gamma$, and write
\begin{equation}
A(\chi,\theta)= {a_0\,\theta\,N\over \tau_0}\,\bigl[\Gamma(s_C,\chi) +\kappa\, \rho(\theta)\bigr],
\end{equation}
where $\kappa$ is a dimensionless parameter, the factor $\theta$ has been inserted for dimensional reasons, and $a_0$ is a dimensionless quantity to be determined as follows.

Separate the right-hand side of Eq. (\ref{chidot}) into parts proportional to $\Gamma$ and $\rho$, and then write this equation in the form
\begin{eqnarray}
\label{chidot1}
\nonumber
{\tau_0\,C_V^{e\!f\!f}\,\dot \chi\over N} &=& \Gamma(s_C,\chi)\,\Bigl[\Lambda^{eq}\,v_0\,s_0 + a_0\,(\theta - \chi)\Bigr]\cr \\ &+& a_0\,\kappa\,\rho(\theta)\,(\theta - \chi).
\end{eqnarray}
In \cite{JSL-MANNING-TEFF-07}, it was argued that athermal ($\rho = 0$)  amorphous systems reach steady state for effective temperatures $\chi$ equal to some function $\hat\chi(q)$, where $q$ is a dimensionless, non-negative measure of the total strain rate.  For time-independent stresses, $q$ is the magnitude of $\tau_0\,D_{in}$. This means that the quantity in square brackets in Eq. (\ref{chidot1}) must vanish at $\chi = \hat\chi(q)$, a condition that we satisfy by setting
\begin{equation}
a_0 = {\Lambda^{eq}\,v_0\,s_0\over \hat\chi(q) - \theta}.
\end{equation}
Thus, Eq. (\ref{chidot1}) becomes
\begin{eqnarray}
\label{chidot2}
&&{\tau_0\,C_V^{e\!f\!f}\,\dot \chi\over v_0\,s_0\,N}= \\
\cr &&{\Lambda^{eq}\over  \hat\chi(q) - \theta} \left[\Gamma(s_C,\chi)\,\bigl(\hat\chi(q) - \chi\bigr) + \kappa\,\rho(\theta)\,(\theta - \chi)\right].\nonumber
\end{eqnarray}

Equation (\ref{chidot2}) is essentially the same $\dot\chi$ equation that we have used in previous applications. The main difference is the prefactor $(\hat\chi - \theta)^{-1}$.  Non-negativity of $a_0$ requires that $\hat\chi(q) > \theta$, which is a plausible and interesting constraint.  The steady-state solution of Eq. (\ref{chidot2}) is
\begin{equation}
\chi_{ss} = {\Gamma(s_C)\,\hat\chi(q) + \rho(\theta)\,\theta\over \Gamma(s_C) + \rho(\theta)}.
\end{equation}
The function $\Gamma(s_C)$ vanishes in the limit of vanishing strain rate $q$; therefore, for fixed, nonzero $\rho(\theta)$,  $\chi_{ss} \to \theta$ as $q \to 0$.  On the other hand, if the strain rate is fixed and $\rho(\theta)$ becomes small, then $\chi_{ss} \to \hat\chi(q)$. As pointed out in \cite{JSL-MANNING-TEFF-07}, the crossover between these limiting behaviors takes place at very small strain rates for small $\rho(\theta)$, and therefore it can be very difficult to determine whether a glass transition has occurred.  At higher temperatures, this crossover occurs at higher strain rates, and the condition $\hat\chi(q) > \theta$  requires that $\hat\chi$ be a function of $\theta$ in some circumstances.  For the moment, we note that physically realistic   systems do not probe the extreme limit of vanishingly small strain rate, and we therefore  assume that  $\hat\chi(q) - \theta \cong \chi_0 - \theta$ is a positive constant for situations in which the system is deforming at experimentally accessible rates.

\section{Summary of STZ Equations}
\label{STZsummary}

We conclude this part of the paper by summarizing the STZ equations in their most usable versions, that is, in the limit in which the relaxation of the STZ variables $\Lambda$ and $m$ is much faster than the rates at which plastic deformation and the effective temperature respond to changes in the external driving forces.  Many of these equations are the same as the ones that appear -- in more general tensorial versions -- in \cite{JSL-STZ-PRE08}.  As noted previously, however, there are some  differences.

The rate of inelastic deformation, given here in Eq. (\ref{Ddef}), is a function of the configurational shear stress $s_C$ (assuming no appreciable contribution from the viscous stress in the kinetic/vibrational subsystem) and the effective temperature $\chi$
\begin{equation}
D_{in}^{dev} = \Lambda^{eq}(\chi)\,f(s_C,\chi),
\end{equation}
where
\begin{equation}
\Lambda^{eq}(\chi) \approx e^{- e_Z/\chi},
\end{equation}
and
\begin{equation}
\nonumber
f(s_C,\chi) = {\epsilon_0\over\tau_0}\,{\cal C}\left({\cal T}-\,m^{eq}\right).
\end{equation}
Here, we have reverted to the earlier notation, $\epsilon_0 = N\,v_0/V$, which is the ratio of a molecular volume $v_0$ associated with STZ transitions to the volume per molecule in the system as a whole, and is of the order of unity. The STZ formation energy $e_Z$ previously was denoted by $k_B\,T_Z$.  In \cite{JSL-STZ-PRE08}, $T_Z$ was found to be larger than the glass temperature by a factor of about $30$ for a metallic glass; and the time constant $\tau_0$ was of the order of a femtosecond. We have abbreviated the functions ${\cal C}$ and ${\cal T}$ as follows
\begin{equation}
{\cal C} = \,R_0(s_C,\chi,\theta)\,\cosh \left({v_0\,s_C\over \chi}\right)
\end{equation}
and
\begin{equation}
{\cal T} = \tanh \left({v_0\,s_C\over \chi}\right).
\end{equation}
$R_0(s_C)$ is an arbitrary, symmetric function of the shear stress $s_C$. $m^{eq}(s_C,\theta)$ is the stationary solution of Eq. (\ref{mdot2})
\begin{eqnarray}
\label{meq}
\nonumber
&&m^{eq}(s_C,\theta)={s_0\over 2\,s_C}\,\left[1+ {s_C\over s_0}\,{\cal T}+{\rho(\theta)\over 2\,{\cal C}}\right]\cr\\ &&- {s_0\over 2\,s_C}\sqrt{\left[1+ {s_C\over s_0}{\cal T}+{\rho(\theta)\over 2{\cal C}}\right]^2 - 4{s_C\over s_0}{\cal T}}.~~~~~~
\end{eqnarray}
The parameter $s_0$ is a stress that can be determined from the low-temperature yield stress (minimum flow stress) $s_y$ via Eq. (\ref{syield})
\begin{equation}
s_y\,\tanh\left({v_0\,s_y\over \chi_0}\right) = s_0,
\end{equation}
where $\chi_0$ is the steady-state value of $\chi$ in the limit of vanishingly small strain rate.

It is useful to look at the equation of motion for $\chi$, Eq. (\ref{chidot2}), in two special cases.  First, consider the parameter range relevant for deformations of ordinary plastic materials such as metallic glasses.  The experience gained from the studies reported in \cite{JSL-STZ-PRE08} and \cite{JSL-MANNING-TEFF-07} suggests, for temperatures not too far above the glass transition, and for strain rates not extremely small, that we can assume that $\hat\chi(q) \approx \chi_0$ remains constant at a value larger than $\theta$, so that the dimensionless quantity $\hat\chi/(\hat\chi - \theta)$ is a slowly varying function of $\theta$ that can be absorbed into other parameters such as the effective heat capacity and $\kappa$.  When this is true, Eq. (\ref{chidot2}) can be written in the form
\begin{eqnarray}
\label{chidot3}
\nonumber
&&\tau_0\,\tilde c_0\,\dot\chi \cong e^{-\,e_Z/\chi}\cr \\ &&\times \left[\Gamma(s_C,\chi)\,\left(1 - {\chi\over\chi_0}\right) + \tilde\kappa\,\rho(\theta)\,\left(1 - {\chi\over\theta}\right)\right],
\end{eqnarray}
where $\tilde c_0$ and $\tilde\kappa$ are dimensionless constants of the order of unity. To use this equation, we need the explicit expression for $\Gamma$
\begin{equation}
\Gamma(s_C,\chi) ={{\cal N}(s_C,\chi)\over 1 - (m^{eq}\,\chi/s_0\,v_0)\,\tanh^{-1}(m^{eq})},
\end{equation}
where
\begin{eqnarray}
\nonumber
&&{\cal N}(s_C,\chi)= \rho(\theta)\, {m^{eq}\chi\over v_0\,s_0} \tanh^{-1}(m^{eq}) \cr \\ \nonumber && + 2\,{\cal C}(s_C)\,\bigl[{\cal T}(s_C) - m^{eq}\bigr]\cr \\&& \times \left[{s_C\over s_0} - {\chi\over v_0\,s_0}\,\tanh^{-1}(m^{eq})\right]  ,
\end{eqnarray}
and $m^{eq}(s_C,\chi)$ is given by Eq. (\ref{meq}).

Second, consider the athermal limit of Eq. (\ref{chidot2}) by setting $\theta = 0$ and $\rho(\theta) = 0$.  In this case, we have
\begin{equation}
\label{chidot4}
\tau_0\,\tilde c_0\,\dot\chi \cong e^{-\,e_Z/\chi}\,\Gamma(s_C,\chi)\,\left(1 - {\chi\over\hat\chi(q)}\right),
\end{equation}\\
where, now, $\tilde c_0 = c_V^{e\!f\!f}/\epsilon_0$, and $c_V^{e\!f\!f}$ is the effective heat capacity per unit volume in units of Boltzmann's constant $k_B$.  This limit is appropriate for granular materials, bubble rafts, and the like, where ordinary thermal fluctuations are irrelevant, and the disorder described by the effective temperature is generated only by externally driven deformation.  Thus, only states with stresses above the yield stress are relevant, and Eq. (\ref{meq}) tells us that $m^{eq} = s_0/s_C$ (exactly).  Moreover, when $s_C \gg s_0$, we have
\begin{equation}
s_0\,\Lambda^{eq}\,\Gamma(s_C,\chi) \approx 2\,s_C\,D_{in},
\end{equation}
so that the noise strength is just proportional to the rate at which inelastic work is done on the system.  We have used $\hat\chi(q)$ on the right-hand side of Eq. (\ref{chidot4}), instead of its small-$q$ limit $\chi_0$, because large values of $q$ are more easily attainable for systems in which the intrinsic relaxation time $\tau_0$ is not microscopically small. As shown in \cite{HAXTON-LIU07}, $\hat\chi(q)$ increases rapidly when $q$ grows to values of the order of unity. Thus the restoring term in Eq. (\ref{chidot4}) becomes small; and the resulting rapid growth of $\chi$ produces localized shear failure. This mechanism was shown in \cite{DAUB-08} to provide a plausible explanation of rapid stress drops and localized failure in earthquake faults.

\section{Concluding Remarks}
\label{conclusions}

We have made many simplifying assumptions in developing this thermomechanical version of the STZ theory.  Some of these assumptions were needed only to simplify the presentation, and seem to have little if any physical importance.  For example, it should not be difficult to rewrite this theory in tensor notation, as in \cite{JSL-STZ-PRE08}, and apply it to spatially nonuniform situations with orientationally varying stress and flow fields.  It will be technically more difficult to deal with situations in which both volumetric and shear deformations are occurring and are coupled to each other; but here again there seems to us to be no problem in principle.

Yet another example of simplification is that, throughout this series of three papers, we have dropped terms that would have described thermoelasticity or, more pertinently in the context of nonequilibrium phenomena, thermo-viscoelasticity.  Here too, we see no intrinsic difficulties.  In fact, we see attractive opportunities to use a thermo-viscoelastic version of this theory for studying the behavior of glasses subject to thermal cycling in the neighborhood of the glass temperature.

One of our more problematic simplifications is our assumption that we can distinguish elastic from plastic strains, and use the elastic strain as an independent argument of thermodynamic functions such as the internal energy or the entropy. As we have stated here and in earlier papers, we maintain that the plastic strain, necessarily measured from some reference configuration (possibly evolving), cannot be a physically meaningful variable for determining the current state of the system or predicting its subsequent motion. Thus, we have insisted on expressing our equations of motion in Eulerian coordinates, and using the internal state variables to carry the memory of recent deformations.

This self-imposed requirement leaves us with an as-yet unsolved problem regarding elasticity. The problem is compounded here by our recognition of the extended thermodynamic roles played by internal degrees of freedom, which, as we have seen, may store energy in recoverable forms as well as relax irreversibly toward states of equilibrium.  In such  situations, it is unclear to us whether ``elastic'' behavior is always the same as ``reversible'' behavior, or whether the conventional Kroner-Lee \cite{KRONER-60,LEE-69} decomposition of elastic and plastic displacements is generally correct. We have evaded these issues so far by restricting our attention to infinitesimally small elastic displacements. However, we suspect that these questions now require more serious attention.

Our list of topics needing further investigation includes the choice of rate factors in the STZ theory.  Our most notable departure from earlier STZ results is the relatively simple, $\chi$-dependent transition rate shown in Eq. (\ref{Rate}). This formula is primarily a result of our statistical interpretation of the second law of thermodynamics in \cite{EB-JSL-09-I}; it is related to the two-temperature theory only in the sense that it is the effective temperature $\chi$, and not the thermal temperature $\theta$, that governs the configurational subsystem's motion toward statistically more probable states.  So far as we can tell, this result does not substantially change previous conclusions, e.g. in \cite{JSL-STZ-PRE08,JSL-MANNING-TEFF-07}. In fact, the stronger stress dependence in Eq. (\ref{Rate}) may be needed in order to understand seismic data \cite{DAUB-09}.

This statistical interpretation of the rate factors is especially difficult for jammed states at low temperatures, where the stress is below the yield stress and $\rho(\theta) = 0$. Our theory predicts that, in this situation, $m = \tanh(v_0\,s_C/\chi)$.  This result makes sense for a glass below its glass transition temperature, where thermal fluctuations still can activate transitions between the states of STZ's even if they cannot create new ones. In this case, we can change the inelastic strain by changing the stress, although reequilibration to a new state of deformation might be very slow.

For a granular material, however, the most we can say is that $m = \tanh(v_0\,s_C/\chi)$ is the statistically most likely average orientation of STZ's at the given values of $s_C$ and $\chi$.  Such a state might be achieved by tapping the system, i.e. by artificially introducing something like thermal noise. But the way in which such a jammed system responds to changing stresses has to do with whether it forms force chains or bridging structures or the like.  Such mechanisms cannot be included in a theory of the kind we are discussing here.  Therefore, when talking about granular materials in Sec.\ref{STZsummary}, we have restricted ourselves to unjammed systems that are undergoing deformation.  More generally, this limitation of the STZ theory emphasizes the need for a more thorough investigation of the limits of validity of this theory and of similarly constructed statistical theories of noncrystalline deformation.

\begin{acknowledgments}
JSL thanks Andrea Liu for useful discussions about earlier versions of this paper.  He acknowledges support from U.S. Department of Energy Grant No. DE-FG03-99ER45762.
\end{acknowledgments}

\end{document}